\documentclass[preprint,superscriptaddress,nofootinbib,showkeys,showpacs,tightenlines,fleqn]{revtex4}
\usepackage{graphicx}
\usepackage{bm}
\usepackage{dcolumn}
\newcommand{\Det}{{\rm Det}}
\newcommand{\Tr}{{\rm Tr}}
\newcommand{\tr}{{\rm tr}}

\newcommand{\be}{\begin{equation}}
\newcommand{\ee}{\end{equation}}
\newcommand{\bea}{\begin{eqnarray}}
\newcommand{\eea}{\end{eqnarray}}
\newcommand{\ba}{\begin{array}{l}}
\newcommand{\ea}{\end{array}}

\newcommand{\lab}[1]{\label{#1}}
\newcommand{\re}[1]{(\ref{#1})}

\begin{document}
\preprint{PNU-NTG-05/2005}
\title{Meson-loop contributions to the quark
  condensate from the instanton vacuum}  

\author{Hyun-Chul Kim}
\email{hchkim@pusan.ac.kr} 
\affiliation{Department of Physics and Nuclear Physics \& Radiation
Technology Institute (NuRI), Pusan National University,
609-735 Busan, Republic of Korea}
\author{M.M. Musakhanov}
\email{musakhanov@nuuz.uzsci.net,yousuf@uzsci.net} 
\affiliation{Theoretical Physics Dept, Uzbekistan National
University, Tashkent 700174, Uzbekistan}
\author{M. Siddikov}
\email{Marat.Siddikov@tp2.ruhr-uni-bochum.de}
\affiliation{Institut f\"ur Theoretische  Physik  II,
Ruhr-Universit\" at Bochum, D-44780 Bochum, Germany}
\affiliation{Theoretical Physics Dept, Uzbekistan National
University, Tashkent 700174, Uzbekistan}
\date{September, 2005}

\begin{abstract}
We investigate the quark condensate of the QCD vacuum within the  
instanton vacuum model.  We calculate the meson-loop 
contributions to the dynamical quark mass and quark condensate  
to ${\cal O}(1/N_c)$-, ${\cal O}(m/N_c)$-, and ${\cal O}((m\ln m)
/N_c)$-order corrections.
We find that the meson (especially pion) loops
provide substantial contributions to the dynamical quark mass and as
a result to the quark condensate.  The results indicate that the $1/N_c$
corrections should be reconsidered in the systematical way.  The
present results are consistent with those from chiral perturbation
theory.    
\end{abstract}
\maketitle

{\bf 1.} The QCD vacuum is known to be one of the most complicated
objects due to perturbative as well as non-perturbative fluctuations.
In particular, the quark and gluon condensates, which are the
lowest dimensional ones, characterize its non-perturbative aspects.    
Moreover, the quark condensate is identified as the order parameter
for spontaneous chiral-symmetry breaking (S$\chi$SB) which plays a
key role in low-energy hadronic phenomena: In the QCD sum rule,
these condensates arise from the operator product expansion and are
related to hadronic observables~\cite{Shifman:1978bx}.  In chiral
perturbation theory ($\chi$PT), the free parameter $B_0$ is introduced
in the mass term of the effective chiral Lagrangian at the leading
order~\cite{Gasser:1983yg}, which measures the strength of the 
quark condensate~\cite{Gell-Mann:1968rz}.
  
The instanton picture allows us to study the QCD vacuum
microscopically.  Since the instanton picture provides a natural
mechanism for $S\chi$SB due to the delocalization of single-instanton
quark zero modes in the instanton medium, the quark condensate
can be evaluated.  The instanton vacuum is validated by the two
parameters: The average instanton size $\rho\sim 0.3\, {\rm fm}$ and
average inter-instanton distance $R\sim 1\, {\rm fm}$.  These
essential numbers were suggested by Shuryak~\cite{Shuryak:1981ff} 
within the instanton liquid model and were derived from
$\Lambda_{\overline{\rm MS}}$ by Diakonov and
Petrov~\cite{Diakonov:1983hh}.   These values were recently  
confirmed by lattice measurements
\cite{Chu:vi,Negele:1998ev,DeGrand:2001tm,Faccioli:2003qz,
Bowman:2004xi}. 

In the present work, we want to investigate the meson-loop
contributions to the quark condensate as well as to the dynamical
quark mass $M$, which are identified as  ${\cal O}(1/N_c)$ 
($N_c$ denotes the number of colors), ${\cal O}(m/N_c)$ and   
${\cal O}((m\,\ln\,m)/N_c)$ order corrections, based on an instanton
liquid model for the QCD
vacuum~\cite{Diakonov:1985eg,Diakonov:1995qy,Diakonov:2002fq}.
The model was later extended by introducing the current quark 
masses~\cite{Musakhanov:1998wp,Musakhanov:2001pc,Musakhanov:vu}.
It was assumed in the model that the $1/N_c$ expansion is the reliable
one and the results were obtained in the leading order in this
expansion.  The "quenched" and a $\delta$-function type instanton size
distribution approximations were employed in the
previous works~\cite{Diakonov:1985eg,Diakonov:1995qy,Diakonov:2002fq, 
Musakhanov:1998wp,Musakhanov:2001pc,Musakhanov:vu}.  The "quenched"
approximation is controlled by the $N_f/N_c$ factor, while the width
of the instanton size distribution leads to a $1/N_c$ corrections.  
However, there is no systematical study of the 
${\cal O}(1/N_c)$ corrections to date~\cite{Shafer0204026ShaferPhysRevD2002}.

On the other hand, it is known that the chiral expansion is working
properly in describing light-quark hadronic physics in low-energy
region, though it is quite nontrivial due to $S\chi$SB.  Many years
ago, Novikov et al.~\cite{Novikov:xj} showed that the  
quark condensate beyond the chiral limit acquires large contribution
from the nonanalytic chiral log term, which is of order
$(m\,\ln\,m)/N_c$, and at a small current quark mass (a few MeV) with
model-independent coefficients fixed. At the typical hadronic scale,
i.e. $1\,{\rm GeV}$, it turns out to be a leading correction, since
$|(\ln\,m)/N_c|\geq 1$.  This chiral log term arises from
pion-loop contributions~\cite{Novikov:xj,Gasser:1983yg}.  Thus, to be 
consistent, we have to consider simultaneously all of the
contributions such as  ${\cal O}(1/N_c)$, ${\cal O}(m)$,  ${\cal O}(m/N_c)$ and  
${\cal O}((m\,\ln\,m)/N_c)$ corrections in order to find the quark
condensate beyond the chiral limit within above-mentioned approximations. 

In the previous work~\cite{Musakhanov:1998wp}, the effective action
with the current quark mass was derived, based on the low--frequency
part of the light quark determinant $\Det_{\rm low}$ obtained by
Refs.~\cite{Lee:sm,Diakonov:1985eg}
(see also the review ~\cite{Schafer:1996wv}).  It was explored that
the smallness of the packing parameter $\pi^2(\rho/R)^4\approx
0.1$ makes it possible to average the determinant over collective
coordinates of instantons with fermionic quasiparticles,
i.e. constituent quarks.  The averaged determinant turns out to be the
light-quark partition function $Z[m]$, with "quenched" approximations
used.  In this framework, the dependence of the quark condensate on
the current quark mass $m$ was investigated but the meson-loop (ML)
contributions were kept intact
~\cite{Musakhanov:2001pc,Musakhanov:vu}.  Moreover, these results are 
valid only for the small $m$.  The derivation for the non--small $m$
was further elaborated in our previous work
~\cite{Musakhanov:2005qn,Kim:2004hd} and we will follow it in the
present work to investigate the ML effects on the dynamical quark
mass as well as on the quark condensate.  

Thus, we aim in the present work at examining the dynamical quark mass
$M$ and quark condensate $\langle\bar qq\rangle$ within the framework
of the instanton vacuum model, assuming "quenched" 
and a $\delta$-function type instanton size
distribution approximations.  
We consider only $u,d$ quarks (flavor SU(2)) and neglect other ones.
Then, we can exactly bosonize the $Z[m]$ to treat   
systematically meson fluctuations and investigate  
the following contributions:  ${\cal O}(1/N_c)$, ${\cal O}(m/N_c)$,
and ${\cal O}((m\,\ln\,m)/N_c)$ orders which come from the ML 
contributions.  

\vspace{0.5cm}
{\bf 2.} It was already investigated how to derive the low-frequency
part of the fermionic determinant $\Det_{\rm low}$ and to average it
over collective coordinates of instantons in Refs.~\cite{Musakhanov:2005qn,Kim:2004hd}.
Thus, we start from the following partition function for $N_f=2$:   
\begin{eqnarray}
Z[\hat{m}] &=& \int d\lambda_+ d\lambda_- D\psi D\psi^{\dagger} \exp
\left[\int d^4 x \sum_{f=1}^{2}\psi_{f}^{\dagger}(i\rlap{/}{\partial}
  \,+\, im_{f})\psi_{f} \right. \cr 
&&\left. + \lambda_+ Y_{2}^+   + \lambda_- Y_{2}^- + N_+
\ln\frac{K}{\lambda_+} + N_- \ln\frac{K}{\lambda_- }\right],
\label{Z}
\end{eqnarray}
where $\lambda_\pm$ are the dynamical coupling constants. $K$ is
introduced to make the logarithm
dimensionless~\cite{Diakonov:1995qy,Musakhanov:1998wp,Musakhanov:vu}.
The $\lambda_\pm$ are determined by the saddle-point calculation.
$Y_{2}^\pm$ denote the 't Hooft-type interaction generated by
instantons~\cite{Diakonov:1997sj}: 
\begin{equation}
Y_{2}^\pm = \frac{1}{N_{c}^2 -1}\int d^4 x
\left[\left(1-\frac{1}{2N_c}\right) \det\, iJ^{\pm}(x) +
  \frac{1}{8N_c} \det\, iJ^{\pm}_{\mu\nu}( x)\right]  
\end{equation}
with
\begin{eqnarray}
J^{\pm}_{fg}(x) &=& \int \frac{d^4 k_f d^4 l_g}{(2\pi )^8}\exp
i(k_f -l_g )x q^{\dagger}_f (k_f )\frac{1\pm \gamma_5}{2}q_g(l_g ),\cr
J^{\pm}_{\mu\nu,fg}( x) &=& \int \frac{d^4 k_f d^4 l_g}{(2\pi
)^8}\exp i(k_f -l_g )x  q^{\dagger}_f (k_f )\frac{1\pm
\gamma_5}{2}\sigma_{\mu\nu}q_g (l_g ).
\end{eqnarray}
with $q (k) = 2\pi \rho F(k )\psi (k)$.  The form factor $F(k)$ is
generated by the fermionic zero modes: 
\begin{equation}
F(k) = -
\frac{d}{dt}[I_0 (t )K_0 (t ) - I_1 (t )K_1 (t )]_{t
=\frac{|k|\rho}{2}} .
\end{equation}
We can neglect the tensor interaction $J^{\pm}_{\mu\nu ,fg}(x)$, since
it leads to the contribution of order ${\cal O}(1/N_{c}^2)$.  
Using the following relations, 
\begin{equation}
q(x)=\int \frac{d^4 k}{(2\pi )^4}\exp (ikx) \, q(k),\,\,\,\,
J^{\pm}_{fg}(x) = q^{\dagger}_f (x)\frac{1\pm \gamma_5}{2}q_g (x ),  
\end{equation}
we get:
\begin{eqnarray}
\label{detJ}
\det\left(\frac{iJ^+ ( x)}{g}\right) + \det\left( \frac{iJ^- (
    x)}{g}\right) 
&=&\frac{1}{8g^2} \left[ -(q^\dagger (x) q (x))^2 - (q^\dagger
    (x)i\gamma_5 {\bm\tau} q(x))^2 \right.\cr
&&\left. +\; (q^\dagger (x){\bm\tau} q(x))^2 + (q^\dagger (x)i\gamma_5
    q(x))^2 \right]
\end{eqnarray}
with the color factor $g^2 = (N_{c}^{2}-1)2N_c/(2N_c -1)$.
In the following, we assume that $N_+=N_-=N/2$ and
$\lambda_\pm=\lambda$.  

Now, we bosonize the quark-quark interaction given in Eq.\re{detJ} by
introducing auxiliary meson fields.  The bosonization can be performed
in an exact manner for $N_f=2$.  The quark fields $q$ and $q^\dagger$
are changed under the $SU(2)$ chiral transformation:
\begin{equation}
\delta q=i\gamma_5 {\bm \tau}\cdot {\bm \alpha} q,\,\,\,\, \delta
q^\dagger =q^\dagger 
i\gamma_5 {\bm \tau}\cdot {\bm \alpha}.  
\end{equation}
The auxiliary meson fields are changed as follows:
\begin{equation}
\delta\sigma = 2{\bm \alpha}\cdot{\bm\phi},\;\;
\delta\bm\sigma=2\eta\bm\alpha,\;\; 
\delta{\bm\phi}=-2{\bm\alpha} \sigma,\;\; 
\delta\eta=-2\bm\alpha \cdot \bm\sigma.  
\end{equation}
Using these transformation properties, we can combine the quark and
meson fields in such a way that chiral symmetry is satisfied:
\begin{equation}
\delta q^\dagger
(\sigma+i\gamma_5\bm\tau\cdot \bm\phi )q =0,\;\;\;  
\delta q^\dagger(\bm\tau\cdot\bm\sigma + i\gamma_5\eta)q =0.  
\end{equation}
Thus, the quark-quark interaction in Eq.(\ref{detJ}) is bosonized as
follows: 
\begin{eqnarray}
\int d^4 x \exp \left[\lambda \left(\det \frac{iJ^+}{g} + \det
\frac{iJ^-}{g} \right)\right] &=&\int D\sigma D\bm\phi D\eta
D\bm\sigma \cr
&&\hspace{-6.5cm}\times \exp\left[\int d^4x \left\{ \frac{\sqrt{\lambda}}{2g}
  q^\dagger i(\sigma +i\gamma_5 \bm\tau\cdot\bm\phi
  +i\bm\tau\cdot\bm\sigma +\gamma_5\eta)q
 -\frac{1}{2} (\sigma^2 + {\bm\phi}^2 + {\bm\sigma}^2 +\eta^2
 )\right\}\right] .  
\end{eqnarray}
Thus, the partition function can be written as
\begin{eqnarray}
\label{partfunction}
Z[m] &=&\int d\lambda D\sigma D\bm\phi D\eta D\bm\sigma
\exp\left[N\ln\frac{K}{\lambda}-N - \frac{1}{2}\int d^4 x
(\sigma^2+{\bm\phi}^2+{\bm\sigma}^2+\eta^2)\right. \cr
&& \left. \;+\; {\rm Sp}
\ln\left(\frac{\rlap{/}{p}+i\hat{m}+i\frac{\sqrt{\lambda}}{2g}( 2\pi   
\rho)^2F^2(p)(\sigma+i\gamma_5{\bm
  \tau}\cdot{\bm\phi}+i\bm\tau\cdot\bm\sigma+\gamma_5\eta)  
}{\rlap{/}{p}+i\hat{m} }\right)\right],  
\end{eqnarray}
where ${\rm Sp}$ denotes the functional trace $\tr\int
d^4x\langle x|(...)|x\rangle$. $\tr$ is the trace over
Dirac spin,  color, and flavor spaces.  We assume isospin 
symmetry $m_u = m_d = m$.  

In order to get the partition function $Z[m]$ 
we have to calculate the integrals over the mesons 
$\sigma,\,\bm\phi,\,\eta,\,\bm\sigma$ and the coupling $\lambda .$
Here the integral over $\lambda$ must be taken in the saddle-point
approximation at the last stage.  

For the beginning we neglect meson fluctuations contributions.
 Then, it is enough to find the common saddle point around
 $\lambda$ and constant $\sigma$ (\mbox{other fields}=0)
 which is defined as
\begin{equation}
\frac{\partial  V[m,\lambda , \sigma ]}{\partial \lambda}=
  \frac{\partial V [m,\lambda , \sigma ]}{\partial
  \sigma}=0, 
\end{equation}
where the  potential is
\begin{equation}
V [m,\lambda , \sigma ] = -N\ln\frac{K}{\lambda}+N +
\frac{1}{2}V \sigma^2 - {\rm Sp}\ln\left(\frac{\rlap{/}{p}
+i(m+M(\lambda,\sigma)F^2(p))}{\rlap{/}{p}+im }\right)  
\end{equation}
with $M(\lambda,\sigma)=\sqrt{\lambda} (2\pi \rho)^2\sigma/2g$.  The
common saddle-point around $\lambda$ and 
$\sigma$ is determined by 
\begin{equation}
N=\frac{1}{2} {\rm Sp}\left( \frac{iM(\lambda,\sigma)F^2(p)}
{\rlap{/}{p}+i(m+M(\lambda,\sigma)F^2(p))}\right)=\frac{1}{2}V\sigma^2 .
\label{saddle}
\end{equation}
$\lambda_0$ and $\sigma_0=\sqrt{2N/V}=\sqrt{2/ R^4}$ stand for the
solutions of Eq.(\ref{saddle}).  It is clear that Eq.(\ref{saddle})
defines $M_0=M(\lambda_0,\sigma_0)$ which is identified as the
dynamical quark mass.  Using  
typical values for the inter-instanton distance and the size of the
instanton: $R^{-1}=200 \,{\rm MeV},\,\,\rho^{-1}=600\,{\rm MeV}$,
respectively, we obtain $\sigma^2_0=2(200\,{\rm MeV})^4$.  In the
chiral limit, we get $M_0\simeq360\,{\rm 
   MeV}$ and $\lambda_{0}\simeq M_{0}^2$.  Note that due to
Eq.~\re{saddle} $M_0$ and $\lambda_0$ are functions of the current 
quark mass $m$: Their dependence on $m$ was investigated already in
Refs.\cite{Musakhanov:2001pc,Musakhanov:vu}. 

\vspace{0.5cm}
{\bf 3.} The quantum fluctuations  being taken into account, the 
potential $V[m,\lambda , \sigma ]$ 
acquires additional ML corrections and turns out to be an effective one
$V_{eff}[m,\lambda , \sigma ]$ ~\cite{Coleman:1973jx}.  Then, the partition function is
represented as 
\begin{equation}
Z[m]=\int d\lambda Z[\lambda, m]=\int d\lambda
\exp(-V_{eff}[m,\lambda,\sigma]). 
\end{equation}
Note that there is an important difference
between this instanton-generated partition function $Z[m]$ and
traditional Nambu--Jona-Lasinio-type models: In the present work, we
need to integrate over the coupling $\lambda$.  In order to find
$Z[\lambda, m]$, we have to take $\sigma$ as a solution of the
following variational equation: 
\begin{equation}
\frac{\delta V_{\rm eff} [m,\sigma,\lambda]}{\delta\sigma}=0. 
\lab{vacuum}
\end{equation}
by which the saddle point on $\lambda$ is determined as follows:
\begin{equation}
\frac{\partial
  V_{\rm eff} [m,\lambda,\sigma(\lambda)]}{\partial\lambda}=0.
\label{saddlelambda}
\end{equation}
Then, Eq.~\re{saddlelambda} finally defines the coupling constant
$\lambda$ and vacuum field $\sigma(\lambda)$.  We denote generically
the meson fluctuations as $\Phi'_i$.  The effective action and 
the corresponding effective potential $V_{\rm eff} $ can be written as follows:
\begin{equation}
S[m,\lambda,\sigma,\Phi']=V[m,\lambda , \sigma
]+S_V[m,\lambda,\sigma,\Phi'] 
\end{equation}
with
\begin{eqnarray}
 S_V[m,\lambda,\sigma,\Phi'] &=& \int
d^4x\frac{1}{2}({\sigma}^{'2}+{\bm\phi}^{'2}+{\bm\sigma}^{'2}+{\eta}^{'2})\cr
&& -\frac{1}{2\sigma^2}{\rm Sp}\left[\frac{iM(\lambda,\sigma)F^2}
{\rlap{/}{p}+i(m+M(\lambda,\sigma)F^2)} (\sigma'+i\gamma_5\bm\tau\cdot\bm\phi'
+i\bm\tau\cdot\bm\sigma'+\gamma_5\eta')\right]^2 ,  
\end{eqnarray}
and
\begin{equation}
V_{\rm eff} [m,\lambda,\sigma]= V[m,\lambda,\sigma]+
 V_{\rm mes}[m,\lambda,\sigma],  
\label{Veff}
\end{equation}
where
 \begin{eqnarray}
\label{loops}
&&V_{\rm mes}[m,\lambda,\sigma]=\frac{1}{2} {\rm
 Sp}\ln\left(\frac{\delta^2
 S_V[m,\lambda,\sigma,\Phi']}{\delta\Phi_i'(x)\delta\Phi_j'(y)}\right)
=\frac{V}{2}\sum_i\int\frac{d^4q}{(2\pi)^4}
\\\nonumber
&\times& \ln \left[1-\tr \frac{1}{\sigma^2}\int\frac{d^4p}{(2\pi)^4} 
\frac{M(\lambda,\sigma)
 F^2(p)}{\rlap{/}{p}+i(m+M(\lambda,\sigma)F^2(p))}\right.
  \left.\Gamma_i\frac{M(\lambda,\sigma)
 F^2(p+q)}{\rlap{/}{p}+\rlap{/}{q}+i(m+M(\lambda,\sigma)F^2(p+q)
 )}\Gamma_i \right].    
 \end{eqnarray}
The summation over $i$ runs over all corresponding meson
fluctuations: $\sigma'$, $\bm\phi'$, $\bm\sigma'$, and $\eta'$.  The
corresponding spin-flavor matrices are given as
$\Gamma_i=(1,\,i\gamma_5\bm\tau,\, i\bm\tau,\,\gamma_5)$.  Certainly,
the quantum fluctuations make the the coupling constant $\lambda$
shifted from $\lambda_0$ to $\lambda_0+\lambda_1$ and $\sigma$ from
$\sigma_0$ to $\sigma_0 + \sigma_1$, where $\lambda_1/\lambda_0$ and 
$\sigma_1/\sigma_0$ are of order ${\cal O}(1/N_c)$. 

We now seek for the common solution of
Eqs.(\ref{vacuum},\ref{saddlelambda}).  First, consider  
Eq.(\ref{saddlelambda}): 
\begin{eqnarray}
\label{dlambda}
\lambda\frac{\partial V_{\rm eff}[m,\lambda,\sigma]}{\partial\lambda}
&=&N-\frac{1}{2}{\rm Sp}\frac{iM(\lambda,\sigma)F^2}
{\rlap{/}{p}+i(m+M(\lambda,\sigma)F^2)}
+\frac{V}{2}\sum_i\int\frac{d^4q}{(2\pi)^4}\left[\sigma^2\right. \cr
&&\hspace{-3cm} \left. -{\rm tr}\int\frac{d^4p}{(2\pi)^4}\frac{M(\lambda,\sigma)
F^2(p)}{\rlap{/}{p}+i(m+M(\lambda,\sigma)F^2(p))}\Gamma_i\frac{M(\lambda,\sigma)
F^2(p+q)}{\rlap{/}{p}+\rlap{/}{q} +i(m+M(\lambda,\sigma)F^2(p+q)
)}\Gamma_i \right]^{-1}\cr 
&&\hspace{-3cm}\times\left[-{\rm tr}\int\frac{d^4p}{(2\pi)^4}
\frac{M(\lambda,\sigma)
  F^2(p)}{\rlap{/}{p}+i(m+M(\lambda,\sigma)F^2(p))}\Gamma_i
\frac{M(\lambda,\sigma) F^2(p+q)}{\rlap{/}{p}+\rlap{/}{q}  
+i(m+M(\lambda,\sigma)F^2(p+q) )}\Gamma_i\right. \cr
&& \hspace{-3cm}\left. + i\,{\rm tr}\int\frac{d^4p}{(2\pi)^4}
\left(\frac{M(\lambda,\sigma)
    F^2(p)}{\rlap{/}{p}+i(m+M(\lambda,\sigma)F^2(p))}\right)^2 
\Gamma_i\frac{M(\lambda,\sigma) F^2(p+q)}{\rlap{/}{p}+\rlap{/}{q}
+i(m+M(\lambda,\sigma)F^2(p+q) )}\Gamma_i \right]\cr
&=& 0.  
\end{eqnarray}
From Eq. \re{vacuum}, we obtain: 
\begin{eqnarray}
\label{dsigma}
&&\sigma\frac{\partial V_{\rm eff}
  [m,\sigma,\lambda]}{\partial\sigma}= V\sigma^2 -{\rm
  Sp}\left[\frac{iM(\lambda,\sigma)F^2} 
{\rlap{/}{p}+i(m+M(\lambda,\sigma)F^2)}\right] 
+\frac{V}{2}\sum_i\int\frac{d^4q}{(2\pi)^4}
\\\nonumber
&&\times\left[\sigma^2-\tr\int\frac{d^4p}{(2\pi)^4}
\frac{M(\lambda,\sigma)
  F^2(p)}{\rlap{/}{p}+i(m+M(\lambda,\sigma)F^2(p))}\right.
\left.\Gamma_i\frac{M(\lambda,\sigma)
    F^2(p+q)}{\rlap{/}{p}+\rlap{/}{q} 
+i(m+M(\lambda,\sigma) F^2(p+q) )}\Gamma_i \right]^{-1}
\\\nonumber
&&\times
\left[ 2 i\,\tr \int\frac{d^4p}{(2\pi)^4}
\left(\frac{M(\lambda,\sigma)
    F^2(p)}{\rlap{/}{p}+i(m+M(\lambda,\sigma)F^2(p))}\right)^2\right.
\left. \Gamma_i\frac{M(\lambda,\sigma)
  F^2(p+q)}{\rlap{/}{p}+\rlap{/}{q} 
+i(m+M(\lambda,\sigma)F^2(p+q) )}\Gamma_i \right] = 0  .
\end{eqnarray}
We assume the applicability of the $1/N_c$ expansion, which  provide
the solution of the Eqs.(\ref{dlambda}, \ref{dsigma}) by iteration 
starting from $\sigma_0,\lambda_0$. 

 To simplify the
expressions, we introduce the following vertices $V_{2i}(q)$, 
$V_{3i}(q)$, and the meson propagators $\Pi_i (q)$ defined as follows:  
\begin{eqnarray}
\label{vertices}
V_{2i}(q) &=&\tr \int\frac{d^4p}{(2\pi)^4}
\left[ \frac{M_0(p)}{\rlap{/}{p}+i\mu_0(p)}\Gamma_i
 \frac{M_0(p+q)}{\rlap{/}{p}+\rlap{/}{q}+i\mu_0(p+q)}\Gamma_i\right]
 \\
V_{3i}(q) &=&\tr \int\frac{d^4p}{(2\pi)^4}\left[
 \left(\frac{M_0(p)}{\rlap{/}{p}+i\mu_0(p)}\right)^2\Gamma_i
 \frac{M_0(p+q)}{\rlap{/}{p}+\rlap{/}{q}+i\mu_0(p+q)}\Gamma_i\right]
 \\
 \label{propagators}
\Pi^{-1}_i (q)&=&\sigma_0^2 - V_{2i}(q),
\end{eqnarray}
where $M_0(p)=M_0F^2(p),\,\,\,\mu_0(p)=m+M_0(p)$ and
$\sigma_0^2=2R^{-4}$.  Using the results of Eqs. (\ref{dlambda},
\ref{dsigma}), we arrive at 
\bea
&&\frac{M_1}{M_0}\left[\frac{2}{R^4}+\frac{1}{V}{\rm
    Sp}\left(\frac{M_0(p)} 
{\rlap{/}{p}+i\mu_0(p)}\right)^2\right]
=\sum_i\int\frac{d^4q}{(2\pi)^4} (iV_{3i}(q)-V_{2i}(q))\Pi_i (q)
\label{M1}
\\
&&\frac{\sigma_1}{\sigma_0} 
=-\frac{R^4}{4}\sum_i\int\frac{d^4q}{(2\pi)^4} V_{2i}(q)\Pi_i (q). 
\label{sigma2}
\eea

It is of great importance to consider the pion
fluctuations  $\bm\phi'$ at a small pion momentum $q$, since
they lead to the famous first-leading nonanalytic term in the chiral
expansion, also known as the chiral log term with the
model-independent coefficient, as already discussed in the NJL model
\cite{Nikolov:1996jj}.  The inverse propagator 
$\Pi^{-1}_{\bm\phi'}(q)$ of the pion at the small $q\sim m_\pi$ turns
out to be 
$
\Pi^{-1}_{\bm\phi'}(q) =  f_{\pi}^2(m_{\pi}^2 + q^2).
$
At the lowest order with respect to $m$, we find the pion decay
constant $f_{\pi}\approx  93\,{\rm MeV}$ and the pion
mass $m_\pi^2\sim m$.  In the chiral limit, the vertices in the RHS of
Eq.\re{M1} at $q=0$ are given by
\begin{equation}
V_{2\bm\phi_i', m=0}(0)=\frac{2}{R^4},\,\,\,\,i V_{3\bm\phi_i',
 m=0}(0) - V_{2\bm\phi_i', m=0}(0) 
 = 8N_c\int\frac{d^4p}{(2\pi)^4}\frac{p^2M_0^2(p)}{(p^2+M_0^2(p))^2},  
\end{equation}
while the trace in the LHS of Eq.(\ref{M1}) turns out to be 
\begin{equation}
\tr \int\frac{d^4p}{(2\pi)^4}\frac{i\rlap{/}{p}
  M_0(p)}{(\rlap{/}{p}+iM_0(p))^2}= -2(i V_{3\bm\phi_i', m=0}(0) -
  V_{2\bm\phi_i', m=0}(0)).    
\end{equation}
Collecting all the factors, we get for small $q\leq \kappa$
the pion fluctuations $\bm\phi'$ as follows:
\begin{eqnarray}
\label{M1-phi-smallq}
\left.\frac{\sigma_1}{\sigma_0}\right|_{\bm\phi'}
&=&\left.\frac{M_1}{M_0}\right|_{\bm\phi'}
=-\frac{3}{2f_\pi^2}\int_0^{\kappa} \frac{d^4q}{(2\pi)^4} 
\frac{1}{m_{\pi}^2+ q^2}
=-\frac{3}{32\pi^2f_\pi^2}\left(\kappa^2
+m_\pi^2\ln\frac{m_\pi^2}{\kappa^2+m_\pi^2}\right),  
\end{eqnarray}
where $\kappa$ denotes the fact that we consider only the small
momenta as mentioned before.  Note that we put $m=0$ everywhere except
for the value of $m_\pi$.  Here, we want to emphasize that the
coefficient in front of $m_\pi^2\ln m_\pi^2$ is
model-independent.

Since the average size of the instanton renders the present model a
normalization scale, one can assume $\kappa=\rho^{-1}$ and
obtain the pion fluctuations as follows: 
\bea
\lab{M1-phi}
\left.\frac{\sigma_1}{\sigma_0}\right|_{\bm\phi'}\approx
\left.\frac{M_1}{M_0}\right|_{\bm\phi'} 
\approx-\frac{3}{32\pi^2f_\pi^2\rho^2}(1
+m_\pi^2\rho^2\ln m_\pi^2\rho^2)\approx -0.34.
\eea
Thus, we find that the pion loops provide the non-analytical log term
$\frac{1}{N_c}m\ln m$ and yield a rather large contribution to the   
${\cal O}(1/N_c)$ corrections.  

\begin{figure}[t]
  \centering
\includegraphics[height=6.5cm]{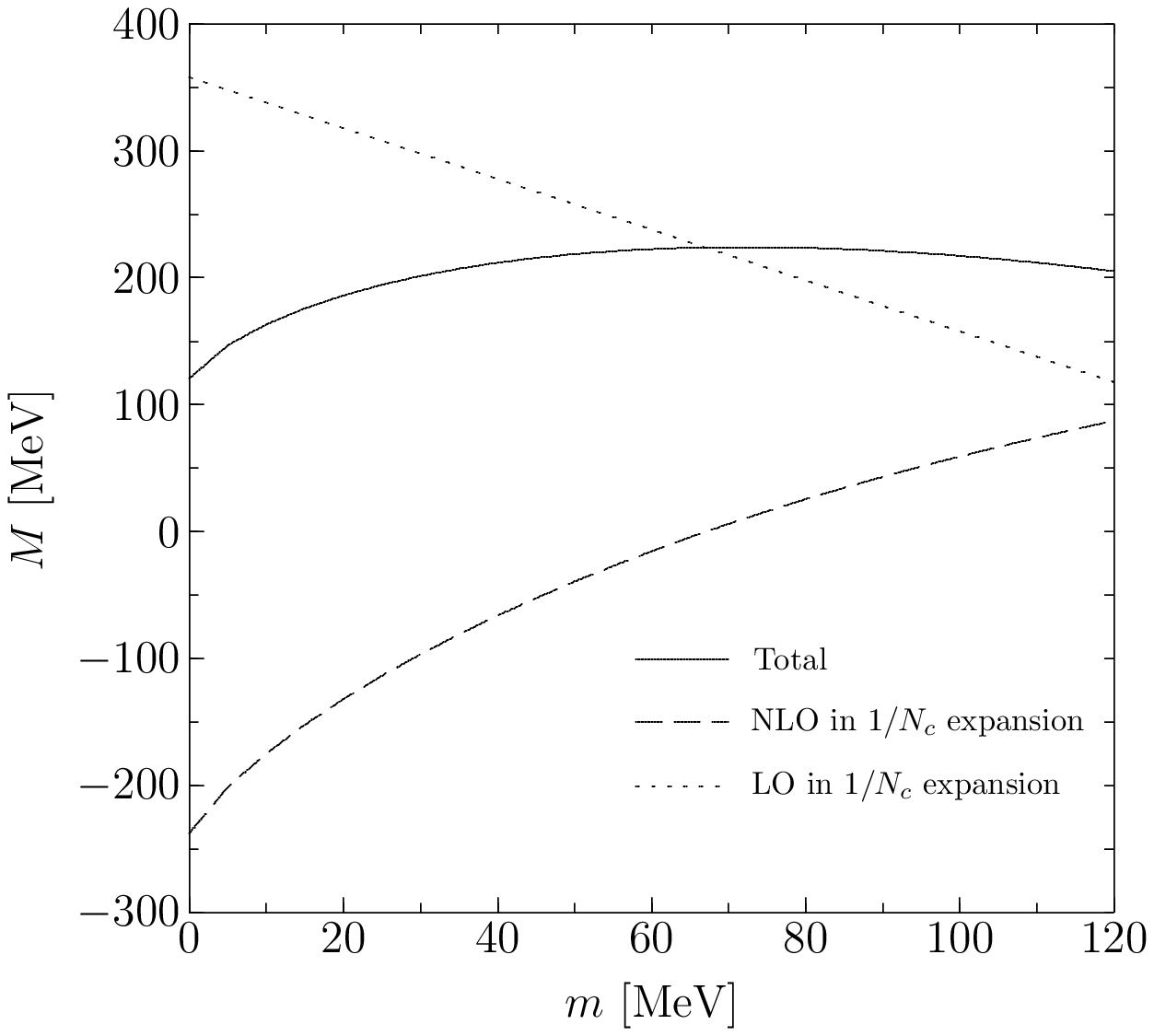}
\hspace{0.5cm}
\includegraphics[height=6.5cm]{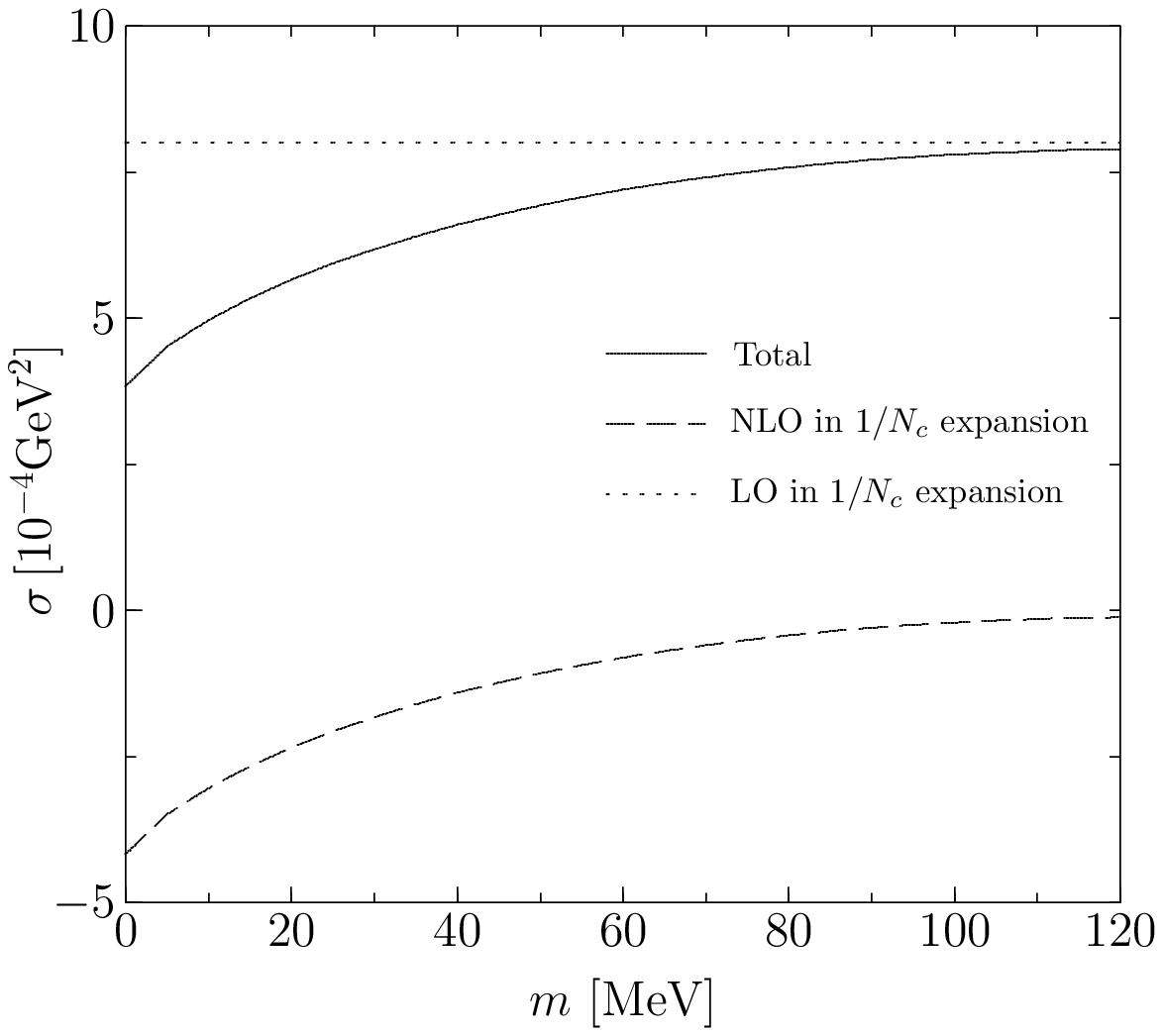}
\caption{$m$ dependence of the dynamical quark mass $M$ and
vacuum average $\sigma$.  The dotted curve depicts the leading order
(LO) in the $1/N_c$ expansion, while the dashed one draws the 
next-to-leading order (NLO) (ML).  The solid curve represents the
total contribution.}
  \label{fig:1}
\end{figure}
Now, we are in a position to calculate the relative shifts $M_1/M_0$ and
$\sigma_1/\sigma_0$.  To simplify the computation, we employ the
dipole-type form factor which is known to produce very similar results
to those obtained with the zero-mode
one~\cite{Petrov:1998kg,Choi:2003cz,Kim:2004hd} and incorporate its 
correct asymptotic behavior above the momentum region $p>2\,{\rm GeV}$: 
\bea
\lab{F(p)}
F(p<2 {\rm GeV})=\frac{\Lambda^2}{\Lambda^2+p^2},\,\,\, F(p>2 {\rm
  GeV})=\frac{\sqrt{2}}{p^3}, 
\eea
where $\Lambda \sim \sqrt{2}/\bar{\rho} \simeq 850 \,{\rm MeV}$.  At
$N_c=3$, we obtain the following results:
\bea
\frac{M_1}{M_0} &=& -0.66 - 4.64 m - 4.01 m\ln m
\lab{formulaM1}
\\
\frac{\sigma_1}{\sigma_0} &=&-0.52 - 4.26\, m - 4.00\, m \ln m,
\lab{formula}
\eea
where $m$ is given in $\rm GeV$.
 The last terms in
Eqs.(\ref{formulaM1},\ref{formula}) exactly correspond to Eq.
\re{M1-phi-smallq}.  

The numerical results for the dynamical quark mass
$M=M_0+M_1$ and vacuum average $\sigma=\sigma_0+\sigma_1$ are depicted in 
Fig.~\ref{fig:1}.  As shown in the left panel of Fig.~\ref{fig:1}, the
meson fluctuations, i.e. the $1/N_c$ corrections, diminish $M$ by
about $70\,\%$ in the chiral limit.  However, as $m$ increases, the $1/N_c$
corrections are logarithmically getting larger.  when $m$ reaches
around $70$ MeV, they turn into positive values.   On the contrary,
the linear $m$ correction decreases linearly, so that it counters with 
the $1/N_c$ ones.  As a result, $M$ increases until $m=70$ MeV and
starts to get reduced rather slowly.  $M$ is pulled down by about
$20\,\%$ at $m\simeq 120\,{\rm MeV}$.  

The right panel of Fig.~\ref{fig:1} shows the dependence of the
 vacuum average $\sigma$ on $m$ due to the meson
fluctuations.  The ML corrections show the similar tendency to the
dynamical quark mass $M$: In the chiral limit, the vacuum is shifted
negatively by around $52\,\%$.  However, ML corrections  
become almost zero at $m\simeq120$ MeV,
$\sigma$ is almost restored to the old vacuum $\sigma_0$ at that
value.   Note that the pion loops contribute
dominantly to $M_1/M_0$ and $\sigma_1/\sigma_0$.  Other mesons do to
them approximately by $10\,\%$.  

\vspace{0.5cm}
{\bf 4.} Now, we proceed to investigate the quark condensate beyond
the chiral limit, based on the above-discussed study, i.e., taking
into account ${\cal O}(m),$ ${\cal O} (1/N_c)$, ${\cal O}(m/N_c )$
and ${\cal O}((m\ln m)/N_c)$ contributions.  The quark condensate can be
easily derived from the partition function: 
\bea
\nonumber
\langle\bar qq\rangle &=&\frac{1}{2V}\frac{d V_{\rm
 eff}[m,\lambda,\sigma]}{d m} =\frac{1}{2V}\frac{\partial
(V [m,\lambda,\sigma]+V_{\rm
  mes}[m,\lambda_0,\sigma_0])}  
{\partial m} 
\\
&=&-\frac{1}{2V}{\rm Sp}\left(\frac{i}{\rlap{/}{p} +i\mu
  (p)}-\frac{i}{\rlap{/}{p}+im}\right) +\frac{1}{2V}\frac{\partial
  V_{\rm mes}[m,\lambda_0,\sigma_0]} 
{\partial m},
\lab{cond}
\eea
where $\lambda=\lambda_0+\lambda_1$, $\sigma=\sigma_0+\sigma_1$, 
$M=M_0+M_1$, and $\mu(p)=m+MF^2(p)$. The first term of Eq.~\re{cond}
can be written as 
\bea
-4N_c\int\frac{d^4p}{(2\pi)^4}\left(\frac{\mu_{0}(p)}{p^2+\mu_{0}^2(p)}
-\frac{m}{p^2+m^2}
+\frac{M_1}{M_0}\frac{M_0(p)(p^2-\mu_{0}^2(p))}{(p^2+\mu_{0}^2(p))^2}\right)
\eea
The second term of Eq.~\re{cond}, which is the contribution from the
direct ML to the quark condensate, is expressed as follows:  
\bea
\nonumber
\frac{1}{2V}\frac{\partial V_{\rm mes}[m,\lambda_0,\sigma_0]}
{\partial m} &=& \frac{i}{2}\sum_i\int\frac{d^4q}{(2\pi)^4}
 \left(\tr\int\frac{d^4p}{(2\pi)^4}
 \frac{M_0(p)}{(\rlap{/}{p}+i\mu_0(p))^2}\Gamma_i
 \frac{M_0(p+q)}{\rlap{/}{p}+\rlap{/}{q}+i\mu_0(p+q)}\Gamma_i\right)
\\
 &&\hspace{-3cm} \times\left(\frac{2N}{V} -
   \tr\int\frac{d^4p}{(2\pi)^4}\frac{M_0(p)}{\rlap{/}{p}+i\mu_0(p)} 
  \Gamma_i
  \frac{M_0(p+q)}{\rlap{/}{p}+\rlap{/}{q}+i\mu_0(p+q)}\Gamma_i\right)^{-1} . 
\eea
If we turn off the meson fluctuations and put $m=0$, we end up with the
well-known formula for the quark condensate in the chiral limit: 
\bea
\langle\bar qq\rangle_{00}=
-4N_c\int\frac{d^4p}{(2\pi)^4}\frac{M_{00}(p)}{p^2+M_{00}^2(p)},
\eea
where $M_{00}\equiv M_{0,m=0}$.
\begin{figure}[b]
  \centering
\includegraphics[height=7cm]{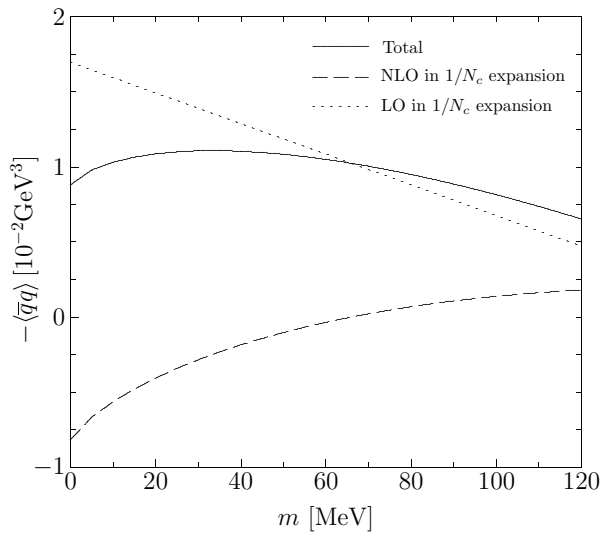}
\caption{$m$ dependence of the quark condensate $\langle\bar
qq\rangle$ (taken with opposite sign).  
The dotted curve depicts the leading order (LO) in the $1/N_c$
expansion, while the dashed one draws the next-to-leading order (NLO)(ML).
  The solid curve represents the total contribution.}
  \label{fig:3}
\end{figure}

Let us first consider the pion fluctuations $\bm\phi'$ for the quark
condensate at small $q$:
\bea
 \frac{1}{2V}\left.\frac{\partial
     V_{\bm\phi'}[m,\lambda_0,\sigma_0]}{\partial
   m}\right|_{q^2\simeq m_\pi^2} =
 12N_c\int\frac{d^4 
p}{(2\pi)^4}\frac{M_0^2(p)\mu_0(p)}{(p^2+\mu_0^2(p))^2}
\int_0^\kappa \frac{d^4 q}{(2\pi)^4 f_{\pi}^2 (m_{\pi}^2+ q^2)}.  
\eea
Note that we keep $m$ only in $m_\pi^2$ as before. Thus, we arrive at  
\begin{equation}
\langle\bar qq\rangle = \langle\bar
qq\rangle_{00}\left(1-\frac{3}{2}\int_0^\kappa \frac{d^4 
q}{(2\pi)^4}\frac{1}{ f_{\pi}^2 (m_{\pi}^2+ q^2)} \right) \;\simeq\;
\langle\bar qq\rangle_{00} \left(1-\frac{3m_\pi^2}{32\pi^2 f_\pi^2}
  \ln m_\pi^2\right),
\label{cond-phi-small-q} 
\end{equation}
where we have utilized the result for $M_1/M_0$ obtained in
Eq.~\re{M1-phi-smallq}.  The second term in Eq.~\re{cond-phi-small-q}
is the eminent analytic 
Refs.~\cite{Gasser:1983yg,Novikov:xj}.   
Having carried out the similar calculation as done for the dynamical quark 
mass $M$ and vacuum average $\sigma$, 
we finally derive the expression for the quark condensate, 
including all ML contributions of  ${\cal O}(1/N_c),$ ${\cal O}(m/N_c)$ and
${\cal O}((m\ln m)/N_c)$ orders:
\begin{equation}
\langle\bar qq\rangle=\langle\bar qq\rangle_{m=0}\left(1 - 18.53\, m -
  7.72\, m \ln m\right), 
\label{cond1}
\end{equation}
where $\langle\bar qq\rangle_{m=0} = 0.52 \langle\bar qq\rangle_{00}$ 
and $m$ is given in $\rm GeV$.
The last term in Eq.~(\ref{cond1}) known as the chiral log term
corresponds to the model-independent expression given in
Eq. \re{cond-phi-small-q}, as it should be.  

Figure~\ref{fig:3} depicts the quark condensate expressed in
Eq.~(\ref{cond1}) (with opposite sign).  The dotted curve draws
the leading order result, while the dashed one represents the ML
corrections.  The solid one represents the quark condensate including
all ML corrections to orders ${\cal O}(1/N_c),$ ${\cal O}(m/N_c)$ and
${\cal O}((m\ln m)/N_c)$.  As shown in Fig.~\ref{fig:3} as well as in 
Eq.(\ref{cond1}), the quark condensate in leading order (without the ML corrections) 
is a monotonically decreasing function of $m$.  The ML 
corrections being included, the quark
condensate is diminished at $m=0$ by approximately $50\,\%$, whereas
at $m=120\,{\rm   MeV}$ it acquires the corrections by around
$10\,\%$.  Thus, taking into account the ML corrections, which
includes the chiral log term, we find that the quark condensate starts
to grow as $m$ increases until $m\sim 40 \,{\rm MeV}$ and then falls
slowly away.  Note that the main ML contribution to ${\cal   O}(1/N_c)$,
${\cal O}(m/N_c)$ and ${\cal   O}((m\ln m)/N_c)$ corrections for
$\langle\bar{qq}\rangle/\langle\bar qq\rangle_{00}$ arises from the
pion loops, while other mesons contribute to ${\cal O} 
(1/N_c)$ and ${\cal O}(m/N_c)$ terms by a few $\%$. 

\vspace{0.5cm}
{\bf 5.} We would like to find also the strange
quark condensate. 
For this purpose, we consider the sum rule in $\chi$PT for the strange
quark condensate~\cite{Gasser:1984gg}.  In $SU(2)$, the current quark
mass is expressed in terms of a diagonal matrix: 
$\hat{m} = (m_u, m_d) = m {\bf 1} + \delta m \tau_3/2$ with 
$m=(m_u+m_d)/2$ and $\delta m= m_1-m_2$.  Here, we are interested in
considering the mass term $\delta m$ which breaks isospin symmetry.
In order to treat it, we introduce an external fields $\bm s$: 
$s_1=s_2=0$ and $s_3=i\delta m/2$.  To find the effect of isospin
breaking in the quark condensate $(\langle\bar uu\rangle-\langle\bar
dd\rangle)/\langle\bar uu\rangle$, we need only to take into account
${\cal O}(\delta m)$ term, neglecting all other corrections. 
 
In the presence of the external field $\bm s$, we expect also the
vacuum field $\bm\sigma$. The effective potential in this case 
can be expressed as follows:
\bea
V_{\rm eff}[\sigma,\bm\sigma, m]\approx  \frac{V}{2}(\sigma^2 +{\bm\sigma}^2)
-\Tr\ln\frac{\rlap{/}{p} + i\bm\tau\cdot\bm s
  +i(m+M(\lambda,\sigma,\bm\sigma)F^2)}{\rlap{/}{p}+im+
  i\bm\tau\cdot\bm s} -N\ln\frac{K}{\lambda}+N,
\eea
where $\lambda, \sigma, \bm \sigma$ are defined by the following
vacuum equations: 
\begin{equation}
\frac{\partial V_{\rm eff}}{\partial\lambda}=0,\,\,\,
\frac{\partial V_{\rm eff}}{\partial\sigma}=0,\,\,\,\,
\frac{\partial V_{\rm eff}}{\partial\sigma_i}=0
\end{equation}
which can be rendered into the following form:
\bea
\frac{1}{2}{\rm Sp} \frac{F^2(p)M_i(m_i+M_iF^2(p)) }{p^2+ (m_i
+M_iF^2(p))^2}=N.
\eea
Here, $M_i=\sqrt{\lambda}( 2\pi \rho)^2(\sigma \pm
\sigma_3)/2g$.  The solution of these equations leads to $\lambda=\lambda
[m,\bm s]$, $\sigma =\sigma[m,\bm s]$ $\sigma_i=\sigma_i[m,\bm s]$.  We
have to put them into $V_{\rm eff}$ and find $V_{\rm eff}=V_{\rm
  eff}[m,\bm s]$. The desired correlation function arises from
\begin{equation}
\left.\frac{\partial V_{\rm eff}[m,\bm s]}{\partial s_3}\right|_{s_3=\delta
  m/2,s_{1,2}=0}=0.  
\end{equation}
We calculate this correlation function within accuracy by
taking into account only  ${\cal O}(\delta m)$ term. 
Hence, the difference of the quark condensates of $u$ and $d$ 
quarks is obtained as follows:
\bea
\langle\bar uu\rangle-\langle\bar dd\rangle =\frac{1}{V}\left[
  {\rm Sp}\left(\frac{-i}{\rlap{/}{p}+i\mu_u(p)}
-\frac{-i}{\rlap{/}{p}+im_u}\right)  
-{\rm Sp}\left(\frac{-i}{\rlap{/}{p}+i\mu_d(p)}
-\frac{-i}{\rlap{/}{p}+im_d}\right)\right].
\eea
We expect that $ | \langle\bar dd\rangle |\, <\, | \langle\bar uu\rangle |$ if
$m_d\,>\,m_u$.  Since typical values of the light quark current masses
\cite{Leutwyler:1996eq} are known to be: $m_u =5.1 \,{\rm MeV}$,
$m_d=9.3\,{\rm MeV}$ at the scale $1\,{\rm GeV}$, which is in fact
close to our scale $\rho^{-1}=0.6\, {\rm GeV} $, we derive the effect
of isospin symmetry breaking in the quark condensate as follows:
\begin{equation}
\frac{\langle\bar uu\rangle-\langle\bar dd\rangle}{\langle\bar
  uu\rangle}= 0.026 .
\label{asym}
\end{equation}
Incorporating the Gasser-Leutwyler sum-rule~\cite{Gasser:1984gg}, 
and using the asymmetry value given in Eq.(\ref{asym}), 
we are able to estimate the strange quark condensate given at $m_s=120
\, {\rm MeV}$ as: 
\begin{equation}
\frac{\langle\bar ss\rangle}{\langle\bar uu\rangle}= 0.43,
\label{eq:ratio}
\end{equation}
which is rather small.  The reason lies in the fact that the effect of
isospin symmetry breaking in Eq.(\ref{asym}) is not at all small.
However, we want to mention that it must be considered only as a crude
estimate, since we have neglected here the possible effects of the
next-to-leading corrections in the chiral expansion.  Moreover, we
have to consider flavor SU(3) in order to calculate the strange quark
condensate correctly.     

\vspace{0.5cm}
{\bf 6.} In the present work, we investigated meson-loop contributions to 
the dynamical quark mass $M$ and the quark condensate.  Since the
instanton generated  quark-quark interactions are nonlocal and contain
the form-factor induced by the quark zero-modes, quark and meson loops
in the present approach turn out to be convergent integrals.  Given
the average size of instantons and their inter-distance, the model has
no free parameter at all apart from the quark current mass, so that
the present calculation is a very constraint one. 

We first examined the  ${\cal O}(1/N_c)$, ${\cal O}(m/N_c)$ and ${\cal
  O}((m\,\ln\,m)/N_c)$ meson-loop corrections to the dynamical quark
mass.  It was found that the meson fluctuations pulled the dynamical
quark mass down by approximately $70\,\%$, while the vacuum average
$\sigma$ was shifted negatively by about $50\,\%$.  As $m$ increases,
the effect of those fluctuations is getting smaller and at $m=120$
MeV, the ratios $M_1/M_0$ and $\sigma_1/\sigma_0$ are diminished by
about $20\,\%$ and $10\,\%$, respectively.  The meson fluctuations
being taken into account, the modulus of the quark condensate 
is pulled down by about $50\%$ in the chiral limit, mainly via the 
dynamical quark mass effect. Direct meson-loop contributions leads to 
rather small changes of the quark condensate by around $10\%$, as was
expected from the $1/N_c$ counting.  However, the ${\cal O}((m\ln
m)/N_c)$ corrections enhance it by approximately $10\,\%$ at $m=120$
MeV.   

In conclusion, ${\cal O}(1/N_c)$, ${\cal O}(m/N_c)$, and ${\cal O}((m\ln m)/N_c)$ 
meson-loop corrections (especially ${\cal O}(1/N_c)$ ) 
are not negligible but substantial.  However, it should be noted that
in the instanton liquid model, there are other $1/N_c$ corrections to
the chiral condensate and the dynamical quark mass, which were not
considered here.  The eminent model-independent
nonanalytic chiral log term in chiral perturbation theory at one-loop
order was exactly reproduced.  The meson fluctuations come dominantly
from the pions, as was expected.  Since they are not at all small, it
is required to revise systematically the $1/N_c$ expansion within 
the instanton liquid model, for example, it is necessary to go beyond
"quenched" and other approximations.  However, we expect that such a
modification would probably change slightly the basic parameters such
as the average size and inter-distance of instantons.  The extension
of the present scheme to other observables such as the 
pion decay constant and pion mass is under way.   

\section*{Acknowledgments}
This work was supported by the Korea Research Foundation Grant funded
by the Korean Government(MOEHRD) (KRF-2005-202-C00102).  M.M. is 
grateful to D. Diakonov for a useful discussion and acknowledges the
partial support by the ``Brain Pool'' Program.  M.S. acknowledges
support of the Graduiertenkolleg 841 of DFG.

\end{document}